\documentclass[universe,article,accept,pdftex,moreauthors]{Definitions/mdpi} 
\usepackage{bm}
\firstpage{1} 
\pubvolume{1}
\issuenum{1}
\articlenumber{0}
\pubyear{2022}
\copyrightyear{2022}
\externaleditor{Academic Editor: Mariusz P. Dąbrowski, Adam Balcerzak, Vincenzo Salzano,  Gonzalo J. Olmo}  
\datereceived{26 January 2022} 
\dateaccepted{19 February 2022} 
\datepublished{} 
\hreflink{https://doi.org/} 

\Title{Testing Screening Mechanisms with Mass Profiles of Galaxy Clusters}

\TitleCitation{Testing Screening Mechanisms with Mass Profiles of Galaxy Clusters}

\Author{{Lorenzo} 
 Pizzuti~\orcidA{}}

\AuthorNames{Lorenzo Pizzuti}

\AuthorCitation{Pizzuti, L.}

\address[1]{%
Osservatorio Astronomico della Regione Autonoma Valle d'Aosta,  Loc. Lignan 39, I-11020 Nus, Italy; pizzuti@oavda.it}

\abstract{We present \textsc{MG-MAMPOSSt}, a license-free code to constrain modified gravity models by reconstructing the mass profile of galaxy clusters with the kinematics of the cluster's member galaxies.  We describe the main features of the code and we show the capability of the method when the kinematic information is combined with lensing data. We discuss  recent results and forecasts on two classes of models currently implemented in the code, characterized by different screening mechanisms, namely, chameleon and Vainshtein screening. We further explore the impact of possible systematics in view of application to the data from upcoming surveys. This proceedings summarizes the results presented at the ALTECOSMOFUN workshop in September  2021}


\begin{document}


.


\section{Introduction}




Scalar--tensor theories of gravity represent one of the most general classes of modified gravity (MG) models viable at cosmological scales, characterized by a quite large number of interesting features (see, e.g., \citep{Cantata21}). This family of theories includes several subclasses, such as the Horndeski (e.g., \citep{Deffayet:2011gz}) and the more general Degenerate Higher-Order Scalar--Tensor (DHOST hereafter) theories (e.g., \citep{Zumalacarregui:2013pma,Amendola:2019laa}).

A new dynamical scalar field propagates an additional fifth force which enhances gravity locally on sufficiently large scales, producing possible detectable imprints on the formation and evolution of cosmological structures. In order to match observational constraints on small scales and dense environments, the fifth force should be further suppressed through a so-called screening mechanism, which relies on the non-linear interaction of the scalar field (e.g., \citep{Koyama:2015oma}). 

In this framework, galaxy clusters constitute a powerful tool to investigate modification of gravity at large scales (see \citep{Cataneo19review} for a recent review); in particular, with the combination of the cluster’s mass profiles derived with lensing and internal kinematic analyses of the hot intra-cluster gas (e.g., \citep{Terukina:2013eqa,Wilcox:2015kna,Sakstein:2016ggl}) and of cluster's member galaxies (e.g., \citep{Pizzuti:2016ouw,Pizzuti:2017diz}), it is possible to constrain departures from General Relativity (GR) in a complementary way with respect to other cosmological and astrophysical observations. 

To this end, we developed \textsc{MG-MAMPOSSt}, a \textsc{FORTRAN} code which determines cluster mass profiles in modified gravity  from the kinematics of cluster members; the program is based upon the \textsc{MAMPOSSt} method of \cite{Mamon01}. While the latter assumes that the gravitational interaction is described by GR, \textsc{MG-MAMPOSSt} explores two popular and quite general classes of MG models, characterized by distinct screening mechanisms: chameleon screening (CS) and Vainshtein screening (VS). 

The code further implements a lensing simulation to investigate the capability of a joint lensing and kinematic analysis to constrain the free parameters of the models. The code has been recently made publicly {available}\endnote{\href{https://github.com/Pizzuti92/MG-MAMPOSSt}{https://github.com/Pizzuti92/MG-MAMPOSSt}. Accessed 19 February 2022, see \citep{Pizzuti22man} for the basic usage of the code.}. 

Here we apply the method over a mock catalogue of dark matter haloes to forecast the constraints on the parameters in the two classes of models cited above, obtainable by current and future kinematic and lensing measurements of galaxy cluster mass profiles. This paper summarizes the results presented at the ALTECOSMOFUN'21 conference and is based upon the works published in 2020, 2021 and 2022.

 In Section \ref{sec:theory}, we provide a brief overview of the MG models implemented, in Section \ref{sec:Method} the main features of \textsc{MG-MAMPOSSt} are presented while in Section \ref{sec:results} we show our results and preliminary outcomes from the application of \textsc{MG-MAMPOSSt} over real data. Finally, a summary and a discussion about systematics are provided in Section \ref{sec:conc}
\section{Theoretical Background}
\label{sec:theory}
In scalar--tensor theories, the screening of the fifth force operates in different ways depending on the non-linear interactions of the field. 


Models which implement the chameleon mechanism{---}typically conformally coupled models such as $f(\mathcal{R})$  gravity, { \citep{Buch01}}---rely on the potential of the field to make the mass of the scalar very large at small scales, such that the fifth force does not propagate. In CS, the usual Newtonian potential is modified by an additional term which depends on the gradients of the scalar field  $\phi$ {(see e.g., \citep{Khoury13,Burrage2018}),}
\begin{equation} \label{eq:pot_chameleon}
\frac
{\text{d}\Phi}{\text{d}r}=\frac{GM(r)}{r^2}+\frac{\mathcal{Q}}{M_\text{P}}\frac{\text{d}\phi}{\text{d}r}.
\end{equation}
where $M(r)$ is the total mass enclosed within a sphere of radius $r$, $M_\text{P}=(8\pi G)^{-1/2}$ is the reduced Planck mass, $G$ is the Newton's constant and $\mathcal{Q}$ is a dimensionless coupling constant\endnote{In the literature the coupling constant is often indicated by $\beta$. However, as $\beta(r)$ also denotes the velocity anisotropy profile in kinematic analyses of galaxy clusters, we adopt $\mathcal{Q}$ for the coupling to avoid confusion.} which can assume different values depending on the specific theory. The case of $f(\mathcal{R})$ gravity,  {where
the Einstein--Hilbert action is modified by adding a general non-linear
function of the Ricci curvature scalar $R$, can be shown to be conformally equivalent to a subclass of chameleon models where $ \mathcal{Q} = 1/\sqrt{6}$ (see e.g., \cite{brax08})}. In the above Equation \eqref{eq:pot_chameleon} spherical symmetry has been assumed.

The explicit expression of the field profile depends 
on the matter density distribution $\rho(r)$.  
Here we use the Navarro--Frenk--White (NFW) profile of \citep{Navarro97} to model the matter density in MG frameworks, which has been shown to provide an overall good description of the total mass profile of cluster-size halos in equilibrium configuration from cosmological simulations and observational data in GR (e.g., \citep{Peirani17}) and in MG (e.g., \citep{Wilcox:2016guw}). Nevertheless, the \textsc{MG-MAMPOSSt} method can be easily extended to other mass models in non-standard frameworks.
The NFW model is fully specified by the scale radius $r_\text{s}$ at which the logarithmic derivative of the profile is equal to $-2$ and the ''virial'' radius $r_{200}$ encloses an overdensity 200 times the critical density of the universe.



Given the NFW model for $\rho(r)$,  {we solve Equation \eqref{eq:pot_chameleon}  to obtain the field profile inside (where field gradients are negligilble) and outside  the source (where the potential is subdominant) by assuming the same analytical approximation of e.g., \citep{Terukina:2013eqa}},

\begin{equation}
\label{eq:field}
\phi(x) =
  \begin{cases}
       \sim 0 &   r < S  \\
     -\displaystyle{\frac{\mathcal{Q}\rho_\text{s}r_\text{s}^2}{M_\text{P}}}\displaystyle{\frac{\ln(1+x)}{x}}-\displaystyle{\frac{C}{x}}+\phi_{\infty} & r > S. 
  \end{cases}\
\end{equation}

{In the} above equation $\rho_\text{s}(r_\text{s},r_{200})$ is the characteristic density of the NFW model, $\phi_{\infty}$ is the background value of the field,  $C$ is an integration constant and $x=r/r_\text{s}$,.
The solutions match at the \emph{screening radius}  $S$. Requiring the continuity of the function and of its first derivative at the matching point, $C$ and $S$ are fully specified in terms of ($r_{200}$, $r_\text{s}$, $\mathcal{Q}$, $\phi_\infty$). In particular, the screening radius depends on $r_{200}^3$. Thus, haloes with different masses exhibit a less/more efficient chameleon mechanism.

In CS, due to the conformal structure of the models, photon propagation is not affected by the fifth force.  {This can be seen by the definition of the lensing potential $\Phi_\text{lens}$ which, at linear order, is given by the sum of the potential $\Phi$ and the relativistic potential $\Psi$. The contribution of the fifth force appears with opposite sign in the two terms and cancels out (see e.g., \citep{Burrage2018}).} This means that lensing observations are sensitive only to the standard Newtonian potential.

In Vainsthein screening, which is a typical feature of DHOST theories, the suppression of the fifth force is achieved by working on higher-order derivatives of the field. In this case the mechanism to recover GR can be partly broken inside a massive object, giving rise to a fifth force which depends on the gradients of the matter distribution (see e.g., \citep{Dima:2017pwp}). Assuming spherical symmetry and an NFW model for the matter density distribution, the Poisson equation associated with the Newtonian potential $\Phi$ (which governs the dynamics of galaxies and gas in clusters) is modified according to {(e.g., \citep{Sakstein:2016ggl}):}
\begin{equation} \label{eq:massdyn}
  \frac{\text{d}\Phi}{\text{d}r} \equiv \frac{G M_{\text{dyn}}}{r^2}=\frac{G}{r^2}\left[ M_{\rm{NFW}}(r)+M_{200}\frac{Y_1}{4}\frac{r^2(r_\text{s}-r)}{(r_\text{s}+r)^3}\times[\ln(1+c)- c/(1+c)]^{-1}\right],   
\end{equation}
where $M_\text{NFW}(r)$ is the NFW mass profile, $M_{200}$ is the mass of a sphere of radius $r_{200}$ enclosing an average density 200 times the critical density of the universe and $c=r_{200}/r_s$ is called concentration. Finally, $Y_1$ is a dimensionless parameter describing the fifth force~coupling.

In this class of models, photon propagation is explicitly modified. In particular, the relativistic potential $\Psi$ receives a contribution form the fifth force, giving rise to an effective lensing mass {(e.g., \citep{Pizzuti2021})}:
\begin{equation} \label{eq:masslens}
M_{\text{lens}}(r) =\frac{r^2}{2G}\left[\frac{\text{d}\Psi}{\text{d}r}+\frac{\text{d}\Phi}{\text{d}r}\right]=M_\text{dyn}+M_2, 
\end{equation}
where
\begin{equation}
   M_2=-\frac{5}{4}Y_2\frac{r^2M_{200}}{[\ln(1+c)-c/(1+c)]}\frac{(r_\text{s}+r)}{(r_\text{s}+r)^{3}}. 
\end{equation}

The coupling $Y_2$ appears only in the relativistic sector, i.e., it can be constrained only by lensing observations.
Current constraints for $Y_1$ are of the order of $10^{-2}$  as obtained from stellar probes (e.g., \citep{Saltas:2018mxc,Sakstein:2018fwz,Saltas:2019ius}), and of $10^{-1}$ at the cosmological level with galaxy clusters (e.g.,~\citep{Haridasu21,Laudato21}). As for $Y_2$, only cosmological bounds are available ($\mathcal{O}(1)$ from \citep{Sakstein:2016ggl} and $\mathcal{O}(0.1)$ from \citep{Laudato21}, which worked on a generalization of the model presented here).

\section{The \textsc{MG-MAMPOSSt} Method}
\label{sec:Method}
\textls[-15]{\textsc{MG-MAMPOSSt} is a code aimed at constraining modified gravity models by analysing the kinematics of cluster member galaxies. Derived from the original \textsc{MAMPOSSt} method\endnote{The public version of \textsc{MAMPOSSt} can be found {at} \href{https://gitlab.com/gmamon/MAMPOSSt}{https://gitlab.com/gmamon/MAMPOSSt}, accessed 19 February 2022.} of \citep{Mamon01}, the program was first presented in a preliminary version in \citep{Pizzuti:2017diz}, and then extended and updated by \citep{Pizzuti2021}. The latest version implements both the CS and VS parametrization of the gravitational potentials described in Section \ref{sec:theory}, based on the NFW model for the matter density profile}.

{The input data-set of the \textsc{MAMPOSSt} and \textsc{MG-MAMPOSSt}  procedure is the projected phase space (p.p.s.) of the member galaxies $(R,v_z)$,} where $R$ is the projected radius from the cluster center and $v_z$ represents the line-of-sight (l.o.s.) velocity, computed in the rest frame of the cluster. Assuming dynamical relaxation (i.e. galaxies are collisonless tracers of the gravitational potential) and a Gaussian modelling of the three-dimensional velocity distribution, {the codes computes orbits of cluster members by solving the (stationary) spherical Jeans' equation} to obtain the velocity dispersion profile along the radial direction $\sigma^2_r$ (see e.g., \cite{MamLok05}):
\begin{equation}
\label{eq:sigmajeans}
\sigma^2_r(r)=\frac{1}{\nu(r)}\int_r^{\infty}{\exp\left[2\int_r^s{\frac{\beta(t)}{t}\text{d}t}\right]\nu(s)\frac{\text{d}\Phi}{\text{d}s}\text{d}s},
\end{equation}
where $\nu(r)$ corresponds to the number density profile of tracers, $\beta \equiv 1-(\sigma_{\theta}^2+\sigma^2_{\phi})/2\sigma^2_r$ is the velocity anisotropy profile and $\Phi$ is the total gravitational potential, which carries information about the nature of the gravitational interaction (e.g., Equations \eqref{eq:pot_chameleon} and \eqref{eq:massdyn}).

Given a parametric expression for the above mentioned quantities, \textsc{MG-MAMPOSSt} performs a Maximum Likelihood estimation of the model parameters using data of the p.p.s. {In particular, the code computes the probability that a member galaxy found at the point $(R_i,v_{\text{z},i})$ in the p.p.s. belongs to the orbits distribution described by the model(s).}
In the most general case, the code can work with six free parameters, namely two mass profile parameters (e.g., $r_{200},\,r_\text{s}$), one parameter for the scaling of the number density profile $r_{\nu}$\endnote{
Note that $r_\nu$ is in general different from $r_\text{s}$ as the distribution of galaxies in clusters may not follow the distribution of the total matter (e.g., \citep{Mamon19}).} , one parameter describing the velocity anisotropy profile $\beta$, and two parameters defining the MG model to be constrained ($\mathcal{Q},\phi_\infty$ for CS and $Y_1,Y_2$ for VS).

The number density $\nu(r)$ can be, in general, excluded from the \textsc{MG-MAMPOSSt} fit as it can be measured directly by analysing the projected distribution in the phase space. In the following, we assume an NFW profile to model the galaxy density profile.
As for the velocity anisotropy,  
six possible parametrizations of $\beta(r)$ are currently available in the code (see the technical manual \citep{Pizzuti22man}). As a case study, here we adopt the Tiret anisotropy model of \citep{Tiret01},
\begin{equation} \label{eq:tiret}
\beta_T(r)=\beta_\infty\frac{r}{1+r_\beta},
\end{equation}  
where $\beta_\infty$ is the velocity anisotropy for $r\to\infty$ and $r_\beta$ is the characteristic radius of $\beta_T(r)$ (anisotropy radius). In the current version of the code, $r_\beta$ is assumed to be equal to the scale radius of the mass profile $r_\text{s}$. Moreover, hereafter we will work with the re-scaled anisotropy parameter $\mathcal{A}_\infty=(1-\beta_\infty)^{-1/2}$.

The parameter space can be explored either by computing the likelihood over a multi-dimensional grid of values, or by performing a Monte Carlo Markov Chain (MCMC) based on a simple Metropolis--Hastings algorithm with a fixed-step Gaussian random walk. The code takes few hours to produce a complete chain of $\sim 10^5$ points.

Another feature introduced in \textsc{MG-MAMPOSSt} is the possibility to simulate additional lensing information to be combined with the likelihood from internal kinematics. This is straightforward in the case of CS where lensing is not affected by the fifth force contribution; the lensing distribution is modeled as a Gaussian $P_\text{lens}(r_\text{s},r_{200})$ for the NFW mass profile parameters $r_{200}$, $r_\text{s}$. The central values, the standard deviations and the correlation defining the distribution can be customized by the user. In the forecast analysis presented in Section \ref{sec:results}, we will assume the lensing Gaussian is centered on the true values of the cluster mass profile's parameters.

The case of VS requires a full lensing simulation, as the lensing mass is explicitly modified in terms of the two fifth-force couplings $Y_1,\,Y_2$. In \textsc{MG-MAMPOSSt}, we implemented a simple weak-lensing simulation which generates a mock reduced tangential shear profile assuming a fiducial NFW model in GR. The log-likelihood is given by
\begin{equation}
    \ln\mathcal{L}_\text{lens}(\bm{\theta}_\text{l})=-\frac{1}{2}\sum_{i=1}^{N_\text{b}}\frac{\left[\langle g_\text{t}(R_i)\rangle-\langle g_{\text{t,vs}}(R_i|\bm{\theta}_\text{l})\rangle\right]^2}{\sigma^2_{\text{l},i}},
\end{equation}
where $\langle g_\text{t}(R_i)\rangle$ is the simulated averaged reduced tangential shear profile at projected position $R_i$, $\langle g_\text{t,vs}(R_i)|\theta_\text{l})\rangle$
is the theoretical profile computed in VS for the set of parameters $\bm{\theta}_\text{l}=(r_\text{s},r_{200}, Y_1, Y_2)$. 

The number of bins is fixed to $N_\text{b} = 10$ (in agreement with current lensing surveys, e.g., \cite{Umetsu16}) in the projected radial range $[0.12\,R_\text{up},2.9\,R_\text{up}]$, where $R_\text{up}$ is the maximum value of the projected radius a galaxy can have to be considered in the \text{MG-MAMPOSSt} fit (see Section \ref{sec:results}).
The uncertainties are given by the quadratic sum of two contributions:
\begin{equation}\label{eq:errorlens}
\sigma^2_{\text{l},i}=\sigma^2_{\text{e},i}+\sigma^2_\text{lss},
\end{equation}
where $\sigma^2_{\text{e},i}=\sigma^2_\text{g}/[\pi(\alpha^2_\text{up}-\alpha^2_\text{low})n_\text{g}]$ is the noise due to the intrinsic ellipticity $\sigma^2_\text{g}$ of the sources lying within an annulus between the angles $\alpha_\text{low}$ and $\alpha_\text{up}$, and $\sigma^2_\text{lss}$ is the uncertainty due to the projected large-scale structure. The average number of source galaxies per $\text{arcmin}^2$ is given by $n_\text{g}$. The central values of the NFW parameters from which the shear profile is derived, as well as the lensing uncertainties and  $ n_\text{g}$ can manually set in the input file of \textsc{MG-MAMPOSSt}.

\section{Results}
\label{sec:results}


\subsection{Synthetic Halo Catalogue}
The mock catalogue of dark matter haloes used to test \textsc{MG-MAMPOSSt} has been produced with the \textsc{ClusterGEN} code (e.g., \citep{Pizzuti:2019wte}), a generator of spherically symmetric, isolated distributions of particles in dynamical equilibrium, characterized by Gaussian 3D velocity distributions. We assume that all the systematics are under control, i.e., particles in each halo follow an NFW distribution and their velocity is assigned by assuming that $\sigma^2_r$ is given by Equation \eqref{eq:sigmajeans}. As for the other velocity dispersion components, $\sigma^2_\theta\equiv\sigma^2_\phi= [1-\beta(r)]\sigma^2_r$. We will  comment on the effect of systematics in Section \ref{sec:conc}. {It is important to point out here that a more rigorous way to compute orbits of a spherical system in dynamical equilibrium is to use six-dimensional distribution functions (e.g., \citep{Kazantzidis04,vasiliev2019}). A new version of \textsc{ClusterGEN}  based on this methodology is currently under development.} 

We considered a total of 20 synthetic massive cluster-size haloes generated in GR\endnote{{In principle \textsc{ClusterGEN} can be used to produce mock clusters adopting different modified gravity setups and matter density distributions. In the exercise presented here we focus only on an NFW profile in GR as our fiducial model.}}, as a reasonable number of relaxed galaxy clusters for which high-quality data could be available from upcoming surveys. All haloes are different realizations of the same NFW distribution with $r_{200}=2.0\,\text{Mpc}$ and $r_\text{s}=0.3\,\text{Mpc}$. For the case of CS, since the true values of the NFW parameters affect the results, we will further discuss how the constraints change when varying the mass of the synthetic cluster.

For each halo we generated two p.p.s., obtained by considering 600 and 100 particles within the radial range $[ 0.05\,\text{Mpc},r_{200} ]$,  as an optimistic expectation of the number of member galaxies' spectroscopic redshifts available from upcoming surveys, although not unrealistic. The bounds of the radial range are set in spite of real observations, to exclude the cluster core where the Brightest Central Galaxy (BCG) dominates the internal dynamics. As for the upper bound, we adopt the conservative limit of $R_\text{up}=r_{200}$ to ensure the validity of the Jeans' equation. Finally, as mentioned in Section \ref{sec:Method}, we consider a Tiret model for the velocity anisotropy profile with $\beta_\infty=0.5$ (i.e. $\mathcal{A}_\infty=1.41$).

\subsection{Vainsthein Screening}
We apply the \textsc{MG-MAMPOSSt} method to the synthetic p.p.s. with additional simulated lensing information, assuming the VS parametrization for the gravitational potentials, Equation \eqref{eq:massdyn} and Equation \eqref{eq:masslens}, to constrain the set of parameters $r_{200},\,r_\text{s},\,\mathcal{A}_\infty,\,Y_1,\,Y_2$.
Assuming uniform uninformative priors on each parameter, we perform an MCMC sampling of the joint likelihood:
\begin{displaymath}
\ln\mathcal{L}_{\text{joint}}=\ln\mathcal{L}_{\text{tot}}(r_\text{s},r_{200},\mathcal{A}_\infty,Y_1)+N_\text{h}\ln\mathcal{L}_\text{lens}(\bm{\theta}_\text{l}),
\end{displaymath}
where 
\begin{equation}\label{eq:liketot}
    \ln\mathcal{L}_{\text{tot}}=\sum_i^{N_\text{h}}\left(\ln\mathcal{L}^{\text{dyn}}_i\right),
\end{equation} 
is the total \textsc{MG-MAMPOSSt} likelihood obtained by combining the information of $N_\text{h}$ phase spaces, with $N_\text{h}=1\dots 20$.

As for the lensing simulation, we have set $n_\text{g}=30 \, \text{arcmin}^{-2}$, as expected for the Wide Survey of the Euclid mission \citep{laureijs11}, $\sigma_\text{lss}=0.005$ and $\sigma_\text{g}=0.3$. 

\begin{figure}[H]
\includegraphics[width=8.5 cm]{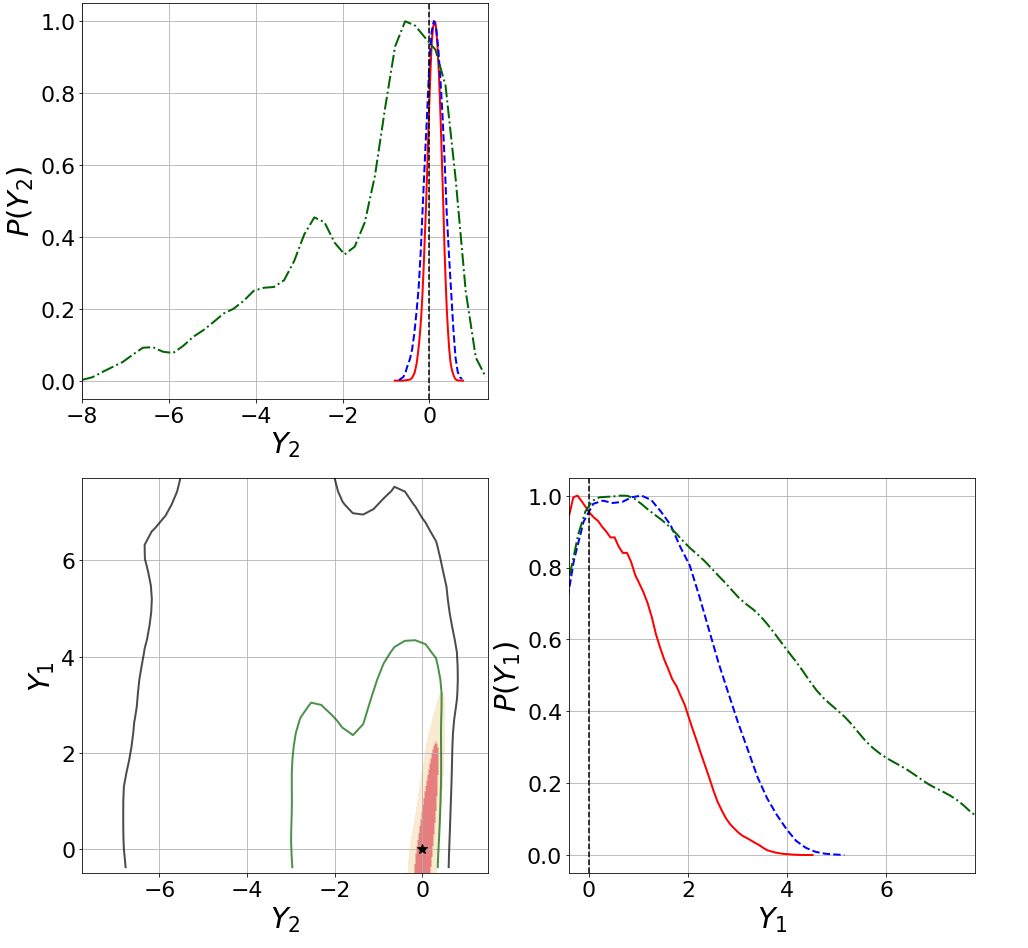}
\caption{\label{fig:lensing1} Results 
 for the \textsc{MG-MAMPOSSt} and lensing forecast in VS for one halo in the sample. {\bf Bottom-left plot}: confidence regions in the space ($Y_1,Y_2$). Dark and light red areas and green and black contours represent the $1\sigma$ and $2\sigma$ regions for the internal kinematic and lensing analysis with 600 tracers and for the lensing simulation only, respectively. {\bf Upper and bottom-right plots}: marginal posteriors of $Y_1$ and $Y_2$. Red lines are for the lensing and kinematics analysis with $N=600$ tracers, blue dashed curves are for the case with $N=100$ tracers and green dash-dotted curves show the lensing-only distributions. The GR expectation (i.e., $Y_1=Y_2=0$) is indicated by the black vertical dashed lines. {From} \citep{Pizzuti2021}.}

\end{figure}

The results of our forecast are summarized in Figure~\ref{fig:lensing1} for one p.p.s. with 600 tracers and in columns two to four of Table~\ref{tab:Y}.  While for $Y_1$ we cannot obtain competitive bounds with respect to stellar probes, even when increasing the number of haloes in the fit, the relativistic coupling $Y_2$ can be constrained at the level of $\mathcal{O}(0.1)$ already with a few clusters, in agreement with the results of \citep{Laudato21} and with a $\sim 2-3$ times improvement with respect to the analysis of \citep{Sakstein:2016ggl}. In particular, for a single halo we obtain  $Y_2=0.08^{+0.32}_{-0.28} $ (600 tracers) and  $Y_2=0.10^{+0.44}_{-0.40} $ (100~tracers). 

\vspace{-12pt}

\begin{table}[H]
\caption[Constraints on $Y_1$ and $Y_2$]{\label{tab:Y} Constraints at 95\% C.L obtained for the parameters of the two modified gravity models presented in this work applying the \textsc{MG-MAMPOSSt} fit with additional lensing information over the mock catalogue of synthetic halos with $r_{200}=2.0\,\text{Mpc}$ and $r_\text{s}=0.3\,\text{Mpc}$. From column two to column six: Vainshtein screening. Column seven and eight: {scalaron field in general chameleon $f(\mathcal{R})$ gravity, related to chameleon field $\phi_\infty$ through Equation \eqref{eq:frcham}}. For both models we show the results when using $\sim$600 and $\sim$100 cluster members in the \textsc{MG-MAMPOSSt} fit.}
\newcolumntype{C}{>{\centering\arraybackslash}X}
\begin{tabularx}{\textwidth}{ccccccc}
\toprule
&  \multicolumn{4}{c}{\textbf{Vainshtein Screening}} &\multicolumn{2}{c}{\boldmath$f(\mathcal{R})$ \textbf{Gravity}}\\
\midrule
& \multicolumn{2}{c}{\boldmath$N=600$} & \multicolumn{2}{c}{\boldmath$N=100$} &  \boldmath$N=600$  & \boldmath$N=100$  \\ 
\midrule

\boldmath$N_\text{h}$ \textbf{Clusters} & \boldmath$Y_1$ & \boldmath$Y_2$ & \boldmath$Y_1$ & \boldmath$Y_2$ & \boldmath$|f_{\mathcal{R}0}|$ & \boldmath$|f_{\mathcal{R}0}|$ \\
\midrule

1 & $ \lesssim 2.75$ & $0.08^{+0.32}_{-0.28} $ &  $ \lesssim 3.56$ &  $0.10^{+0.44}_{-0.40} $ & -- & -- \\

5 & $ \lesssim 1.65$ & $0.06^{+0.20}_{-0.18}$ &  $\lesssim 1.87$ & $-0.08^{+0.31}_{-0.20}$ & $\lesssim 3.37\times 10^{-5}$ & $\lesssim 5.13\times 10^{-5}$\\ 

10 & $ \lesssim 1.24$ & $-0.05^{+0.17}_{-0.13}$ & $\lesssim 1.65$ & $0.01^{+0.24}_{-0.17}$ & $\lesssim 1.12\times 10^{-5}$ & $\lesssim 3.24\times 10^{-5}$\\

15 &  $ 0.04^{+1.00}_{-0.39}$ & $0.01^{+0.12}_{-0.09}$ & $\lesssim 1.20$ & $-0.01^{+0.19}_{-0.16}$ & $\lesssim 9.51\times 10^{-6}$ & $\lesssim 2.43\times 10^{-5}$\\ 

20 &  $0.08^{+0.77}_{-0.34}$ & $0.01^{+0.09}_{-0.08}$ & $\lesssim 1.02$ & $0.01^{+0.16}_{-0.14}$ & $\lesssim 7.11\times 10^{-6}$ & $\lesssim 1.79\times 10^{-5}$\\

\bottomrule

\end{tabularx}
\end{table}

As a preliminary application of \textsc{MG-MAMPOSSt} over real data, we further use the high-precision kinematic and lensing information for the massive relaxed cluster MACS J1206.2-0847 (MACS 1206 hereafter) at redshift $z=0.44$, analysed within the Cluster Lensing and Supernova Survey with Hubble (CLASH, \citep{Postman2012}) and CLASH-VLT~\citep{Rosati2014} collaborations, to constrain $Y_1$ and $Y_2$ in VS (see \citep{Pizzuti2022a} for details). In this case, we have combined the Jeans' analysis of  $375$ cluster member galaxies in the p.p.s. with the strong plus weak lensing data of~\citep{Umetsu16}. The results of 
\begin{equation} \label{eq:constraintsMACS}
Y_1<4.85\,(\text{stat}) \pm 0.2 (\text{syst})\,\;    Y_2=-0.12^{+0.66}_{-0.67}\, (\text{stat})\, \pm 0.21\, (\text{syst}), \end{equation}    
are in very good agreement with the prediction of our forecast analysis. In the above equation, the systematic uncertainties encapsulate the effect of changing the parametrization of the velocity anisotropy profile and the value of the scale radius of the number density profile $r_\nu$ which, as mentioned above, is given by an external fit of the p.p.s.

\subsection{Chameleon Screening}
\begin{figure}[H]
\includegraphics[width=\textwidth]{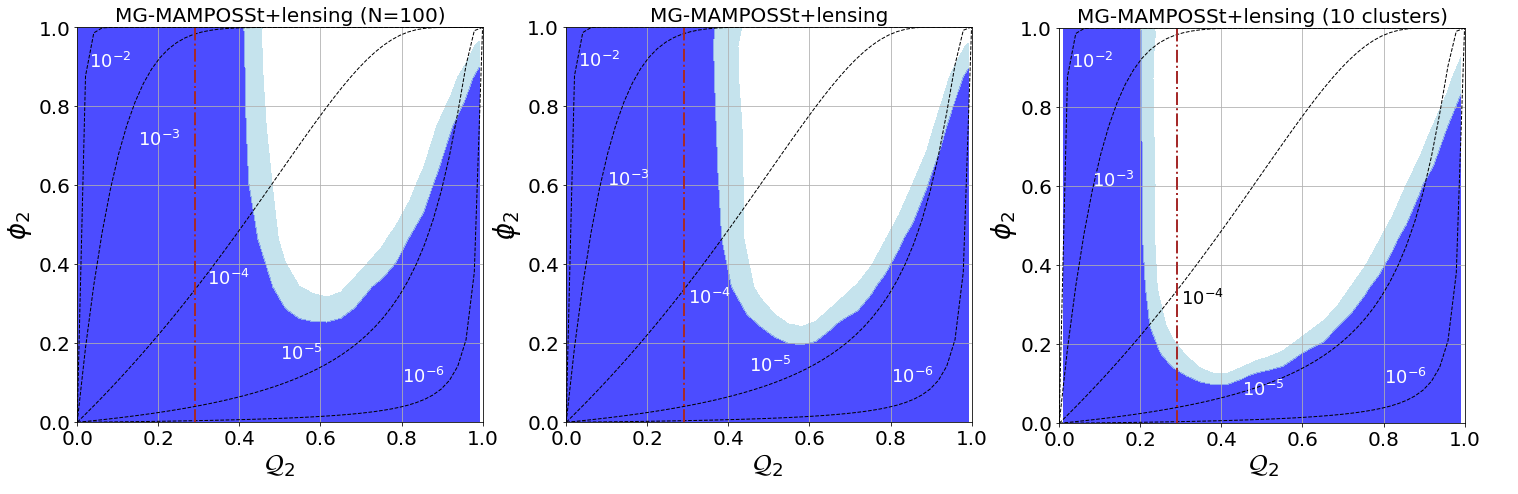}
\caption{\label{fig:cham_ref1} Results for the joint lensing and internal kinematics analysis of the synthetic haloes in our sample. 
The shaded areas  show the allowed regions in the plane $(\mathcal{Q}_2,\phi_2)$ at $3\sigma$ (outer region) and $2\sigma$ (innermost region).  {\bf Left panel}: 100 tracers in the \textsc{MG-MAMPOSSt} fit. {\bf Central panel}: 600 tracers in the \textsc{MG-MAMPOSSt} fit. {\bf Right panel}: combination of 10 clusters, 600 tracers considered in each halo. The black dashed curves are lines of constant $\phi/\mathcal{Q}$ while the vertical brown dash-dotted lines indicate value of the coupling $\mathcal{Q}=1/\sqrt{6}$ corresponding to $f(\mathcal{R})$ gravity. From \citep{Pizzuti2021}.}
\end{figure}

In this case, the MG parameters defining the model are $\phi_\infty$ and $\mathcal{Q}$.
As for VS, we perform an MCMC sampling of the joint kinematic and lensing likelihood. In the Gaussian distribution $P_\text{lens}(r_{200},r_\text{s})$, we assume average realistic uncertainties for current lensing surveys $\sigma_{r_{200}}/r_{200}=0.1$ and $\sigma_{r_\text{s}}/r_\text{s}=0.3$, with a correlation $\rho=0.5$. 
As in e.g. \citep{Terukina:2013eqa,Wilcox:2015kna}, we work with the rescaled variables 
\begin{equation}\label{eq:scaledvar}
\mathcal{Q}_2=\frac{\mathcal{Q}}{1+\mathcal{Q}},\,\,\,\,\,\,\, \phi_2=1-\exp\left[\frac{-\,\phi_{\infty}}{(M_\text{P}c^2_l 10^{-4})}\right],
\end{equation} 
which spawn the range $[0,1]$. In Figure~\ref{fig:cham_ref1} we have plotted the resulting $2\sigma$ (darker areas) and $3\sigma$ (lighter areas) allowed regions in the space ($\phi_2,\,\mathcal{Q}_2$) for three relevant cases.

These results are very similar to that of \citep{Terukina:2013eqa,Wilcox:2015kna} who performed joint X-rays and weak lensing analyses of galaxy clusters, but in the case of \textsc{MG-MAMPOSSt} plus lensing, the excluded region is shifted towards larger values of $\mathcal{Q}_2$. This difference is related to the fact that hot X-ray-emitting gases in clusters and member galaxies are subject to distinct physical processes, despite the fact that they perceive the same gravitational potential. Note that the constraints show only a mild dependence on the number of tracers in the p.p.s., as the strength of the results relies on the additional (lensing) information on $r_\text{s}$ and $r_{200}$, needed to break the statistical degeneracy between the mass profile parameters and the MG parameters in internal kinematic analysis. When increasing the number of haloes in the fit (left panel in Fig. \ref{fig:cham_ref1}, the bounds on the allowed region are noticeably tightened.

As a last step, we focus on the popular $f(\mathcal{R})$ gravity subclass of chameleon models, where $\mathcal{Q}=1/\sqrt{6}$. {In these framework, the fifth force is carried out by the derivative of the function with respect to the Ricci scalar, $f_{\mathcal{R}}=\partial f/\partial\mathcal{R}$, which acts as a scalar field generally known as scalaron. The present-day background value of the scalaron $f_{\mathcal{R}0}$ is connected to the chameleon field $\phi_\infty$ through (e.g., \citep{brax08,Wilcox:2015kna})}:
\begin{equation}\label{eq:frcham}
\phi_\infty=-\sqrt{\frac{3}{2}}\ln(1+f_{\mathcal{R}0})M_\text{P}c^2_l\,
\end{equation}
where $c_l$ is the speed of light. We perform the same kinematic and lensing analysis as for the general chameleon case described above, to constrain the value of the background scalaron field. {It is worth pointing out that the constraints on $f(\mathcal{R})$ gravity derived with this approach are independent of the choice of the function $f$ (see  \citep{Terukina:2013eqa,Pizzuti2021}).} The marginalized distributions of $\phi_\infty$ obtained in different cases are shown in Figure 13 of \citep{Pizzuti2021}
for the combined analysis of 10 haloes. We further list the constraints for $f_{\mathcal{R}0}$ at 95\% C.L. in the last two columns of Table~\ref{tab:Y} when varying the number of haloes considered in the fit. As in CS framework, the screening radius depends on the mass and the density of the sources (i.e., on $r_\text{s}$ and $r_{200}$), we have also generated p.p.s. from a less massive halo with a smaller concentration, characterized by $r_{200}=1.41\,\text{Mpc}$ and $r_\text{s}=0.33\, \text{Mpc}$. The marginalized 10-halo distribution is shown by the blue curve in Figure 13 of citep{Pizzuti2021} 
What we found is that, as expected, smaller clusters exhibit a less efficient screening mechanism, allowing us to better constrain possible departures from GR.
%

As a final result of our forecast, we claim  $|f_{\mathcal{R}0}|\lesssim 1.12\times 10^{-5}$ and $|f_{\mathcal{R}0}|\lesssim 7.11 \times 10^{-6} $ at 95\% C.L. from the joint lensing and internal kinematics analysis of 10 clusters and 20 clusters, respectively, when 600 tracers are considered in the \textsc{MG-MAMPOSSt} fit and $\sigma_{r_{200}}/r_{200}=0.1$, $\sigma_{r_\text{s}}/r_\text{s}=0.3$ in the lensing distribution. For the less massive haloes, we obtain $|f_{\mathcal{R}0}|\le 5.40\times 10^{-6}$ and $|f_{\mathcal{R}0}|\le 3.56\times 10^{-6}$ for 10 and 20 clusters.

\section{Discussion and Conclusions}
\label{sec:conc}
We have introduced the main features of the license-free code \textsc{MG-MAMPOSSt}, developed to constrain modified gravity models with the internal kinematic analysis of galaxy clusters, further equipped with simple lensing simulations for the two classes of scalar tensor theories currently implemented, chameleon screening and Vainsthein screening. 
The tests performed over mock haloes show that the combination of internal kinematic and lensing analyses provides promising results, even for a limited number of clusters. In particular, the relativistic coupling $Y_2$ in Vainsthein screening can be constrained with uncertainties of a few percent with a dozen clusters, assuming reliable high-quality kinematics and lensing information. The exercise has been pushed forward by using the currently available data of the massive galaxy cluster MACS 1206; in this case, we obtain $Y_2=-0.12^{+0.66}_{-0.67}$ (see \citep{Pizzuti2022a}), in agreement with GR prediction and comparable to other constraints at cluster scales for these models \citep{Sakstein:2016ggl,Laudato21}.

In a similar fashion, the allowed region in the parameter space of chameleon gravity ($\phi_2,\mathcal{Q}_2$) can be strongly tightened with joint lensing and internal kinematic analyses of a few galaxy clusters, in a way which is complementary to cluster mass determinations using X-ray analyses of the hot intra-cluster gas (e.g., \citep{Terukina:2013eqa,Wilcox:2015kna}). This indicates that the combination of kinematics of member galaxies, X-rays and lensing probes could provide very strong constraints on the behavior of gravity at cluster scales. 

So far, we have investigated the constraining power of the \textsc{MG-MAMPOSSt} method, assuming that all the systematics are under control. However, deviations from the dynamical relaxation assumption, crucial for the Jeans' analysis in clusters, as well as departures from spherical symmetry, can in principle affect the bounds on the model parameters, giving rise to spurious detection of modified gravity. Calibration and estimation of systematic effects is fundamental in view of the large amount of imaging and spectroscopic data that will be available in the near future. In \citep{Pizzuti19b}, we showed that the analysis of cluster-size dark matter haloes in a $\Lambda$CDM universe indicates departures from GR in $\sim$70\% of cases. Nevertheless, we also found that this percentage can be drastically decreased by looking at the distribution of member galaxies in the p.p.s and the line-of-sight velocity distribution $P(v_z)$. In particular, deviations from Gaussianity in $P(v_z)$, quantified by the so-called Anderson--Darling coefficient $A^2$ \citep{anderson1952}, are strongly related to the probability of finding suitable clusters for the application of our method. This is the case of MACS 1206, characterized by a small value of $A^2\sim 0.6$, in agreement with a nearly Gaussian velocity~distribution.

{Given the current capabilities and limitations of our method, several improvements can be made, which will be available within the next versions of the \textsc{MG-MAMPOSSt} code. In particular, gravitationally bounded structures in some modified gravity theories may not be correctly described by an NFW profile (see, e.g., \citep{Corasaniti20} and references therein). As such, we are planning to extend the parametrizations of the modified gravitational potentials to other mass models which may provide a better description of haloes in MG frameworks. Moreover, the efficiency of the screening mechanism depends on the shape of the matter distribution (e.g., \citep{Burrage19}); thus, one should go beyond the assumption of spherical symmetry in computing the orbits with the Jeans' equation, in order to fully characterize the dynamics in  these MG models.
Finally, our forecast analysis has been performed on synthetic phase spaces generated in GR to explore the constraining power of \textsc{MG-MAMPOSSt}. More realistic constraints can be obtained by computing orbits of member galaxies directly in a modified gravity scenario and comparing the results to real observational data. This can be conducted with the new version of the \textsc{ClusterGEN} code we are developing, which will be made publicly available in the future together with \textsc{MG-MAMPOSSt}.}


\vspace{6pt} 




\funding{LP is partially supported by a 2019 ``Research and Education'' grant from Fondazione CRT. The OAVdA is managed by the Fondazione Cle\'ment Fillietroz-ONLUS, which is supported by the Regional Government of the Aosta Valley, the Town Municipality of Nus and the 
Unite\' des Communes valdotaines Mont-E\'milius.}

\institutionalreview{{Not applicable}}

\informedconsent{{Not applicable}}


\dataavailability{The \textsc{MG-MAMPOSSt} code is publicly available. A test p.p.s. data-set from a synthetic halo is shipped with the code. The data of MACS 1206 are partially provided by~\citep{Umetsu16,Biviano01} by permission. Data will be shared on reasonable request to the corresponding author
with the permission of CLASH-VLT collaboration: \citep{Rosati2014}.} 

\acknowledgments{L.P. acknowledges all the co-authors of the papers presented here - I.D. Saltas, L. Amendola, B. Sartoris, S. Borgani, A. Biviano and K. Umetsu---for their precious contribution.}

\conflictsofinterest{The author declare no conflicts of interest.} 

\newpage
\begin{adjustwidth}{-\extralength}{0cm}
\nointerlineskip\leavevmode
\printendnotes[custom]
\reftitle{References}

\end{adjustwidth}
\end{document}